\documentclass[a4paper,conference]{IEEEtran} 
\ifCLASSINFOpdf
\else
\fi

\usepackage[cmex10]{amsmath}
\usepackage{amssymb,verbatim}
\usepackage{graphicx}

\begin{document}
%
\title{%
Coupling Lemma and Its Application to The Security Analysis of Quantum Key Distribution
}

\author{
\IEEEauthorblockN{Kentaro Kato\\}
\IEEEauthorblockA{
Quantum Communication Research Center\\
Quantum ICT Research Institute, Tamagawa University\\
6-1-1 Tamagawa-gakuen, Machida, Tokyo 194-8610 Japan\\
{\footnotesize\tt E-mail: kkatop@lab.tamagawa.ac.jp} \vspace*{-2.64ex}}
}

\maketitle

\begin{abstract}
It is known that the coupling lemma \cite{A83} provides a useful tool in the study of probability theory and its related areas; it describes the relation between the variational distance of two probability distributions and the probability that outcomes from the two random experiments associated with each distribution are not identical.
In this paper, the failure probability interpretation problem that has been presented by Yuen and Hirota is discussed from the viewpoint of the application of the coupling lemma. 
First, we introduce the coupling lemma, and investigate properties of it.
Next, it is shown that the claims for this problem in the literatures \cite{Y09,Y10} are justified by using the coupling lemma.
Consequently, we see that the failure probability interpretation is not adequate in the security analysis of quantum key distribution.
\end{abstract}

%
\IEEEpeerreviewmaketitle
\section{Introduction}
In the theoretical studies on quantum key distribution (QKD) done in the last decade, 
the trace distance criterion is widely used for evaluating the QKD system for one-time pad ({\it e.g.} \cite{RK05,KRBM07,Ren08,SR08,MR09,SBCDLP09,TLGR12,PR14}). 
In the trace distance criterion, the security notion ``$\varepsilon$-secure" is defined by
\begin{equation}
	d=\frac{1}{2}
	\|\hat{\boldsymbol{\rho}}_{KE}-\hat{\rho}_U\otimes\hat{\rho}_{E}\|
	\leq \varepsilon,
\end{equation}
where $\hat{\boldsymbol{\rho}}_{KE}$ corresponds to the {\it real} system of the key and environment, and $\hat{\rho}_U\otimes\hat{\rho}_{E}$ the {\it ideal} one. 
Further, it is claimed that the parameter $\varepsilon$ is interpreted as the so-called (maximal) failure probability of the QKD protocol.
For this failure probability interpretation of the parameter $\varepsilon$, Yuen and Hirota have respectively presented their objection, together with various other theoretical lack in the security analysis of QKD ({\it e.g.} \cite{Y09,Y10,Y13b,Y13c,Y13d,Hirota12,Hirota13,Hirota14}). 
In this paper, we treat this failure probability interpretation problem.

Tracking back through the development history of the trace distance criterion to seek the origin of the failure probability interpretation, one can arrive at Lemma 1 of the literature \cite{RK05}: ``{\it Let $P$ and $Q$ be two probability distributions. Then there exists a joint probability distribution $P_{XX^\prime}$ such that $P_X=P$, $P_{X^\prime}=Q$, and 
${\rm Pr}_{(x,x^\prime)\gets P_{XX^\prime}}[x\neq x^\prime]=\delta(P,Q)$}", where $\delta(P,Q)$ is the variational distance between $P$ and $Q$. Based on this lemma, they give the failure probability interpretation to $\varepsilon$. However, Yuen has repeatedly claimed that a ``for every" statement would be needed rather than the ``there exists" statement if it justifies the interpretation, and basically, there is no cause one must choose such favorable distribution other than the independent distribution $P_{XX^\prime}=P\cdot Q$ (\cite{Y09,Y10} and its subsequent papers). But, despite the series of criticisms, the failure probability interpretation is still kept \cite{PR14}. This situation motivates us to discuss on the failure probability interpretation problem. 

To discuss the failure probability interpretation problem from a new point of view, we employ
a lemma called the coupling lemma, which was given by Aldous \cite{A83}.
This lemma describes the relation between the variational distance of two probability distributions and the probability that outcomes from the two random experiments associated with each distribution are not identical. 
As we will show later, the use of the coupling lemma would provide a clear perspective on this problem.

\section{Coupling Lemma}
We first introduce the coupling lemma, together with the definition of the variational distance of two probability distributions.
The original statement of the coupling lemma is found in the literature \cite{A83} (Lemma 3.6, p.249). 

\vspace{1em}
\noindent
{\it Definition 1} (variational distance):
Let $X$ and $Y$ be random variables that take on values from a finite alphabet $\mathcal{A}=\{a_1,a_2,\dots,a_N\}$. Let $P_X$ and $P_Y$ denote probability distributions of $X$ and $Y$, respectively. The variational distance between $P_X$ and $P_Y$ is defined by
\begin{equation}
	v(P_X,P_Y)
	=
	\max_{\mathcal{S}\subseteq\mathcal{A}}
	\Bigl[
	P_X(\mathcal{S})-P_Y(\mathcal{S})
	\Bigr],
\end{equation}
or equivalently, by
\begin{equation}
	v(P_X,P_Y)
	=
	\frac{1}{2}
	\sum_{a\in\mathcal{A}}
	\Bigl|
	P_X(a)-P_Y(a)
	\Bigr|.
\end{equation}
\hfill $\blacksquare$

\vspace{1em}
\noindent
{\it Definition 2} (coupling):
There are two probability distributions $P_X$ and $P_Y$ that are defined on a finite set $\mathcal{A}$. Consider a joint probability distribution $P_{XY}$ on $\mathcal{A}^2$ whose marginal distributions are $P_X$ and $P_Y$. We call this $P_{XY}$ a coupling of $P_X$ and $P_Y$.
\hfill $\blacksquare$

\vspace{1em}
\noindent
{\it Theorem 3} (coupling lemma \cite{A83}):
Suppose that $P_X$ and $P_Y$ are given.
\begin{itemize}
	\item[(a)] 
	{\it For every} coupling $P_{XY}$ of
	$P_X$ and $P_Y$, 
	\begin{equation}
		v(P_X,P_Y)
		\leq
		\Pr\{x\neq y\}.
		\label{couplinglemma:case:a}
	\end{equation}
	\item[(b)] 
	{\it There exists} a coupling $P_{XY}$ such that
	\begin{equation}
		v(P_X,P_Y)
		=
		\Pr\{x\neq y\}.
		\label{couplinglemma:case:b}
	\end{equation}
\end{itemize}
\hfill$\square$

\noindent
{\it Proof}: 
One would be able to find the coupling lemma and its proof in the textbooks on probability theory and its applications ({\it e.g.} \cite{L92,T00,LPW09}).
The following proof is obtained by modifying the proof of Theorem A.6 of the literature \cite{PR14}.

(a) Let $P_{XY}$ be an arbitrarily chosen coupling of $P_X$ and $P_Y$. 
It is clear that
$$
	P_{XY}(a,a)\leq P_X(a)
	\quad{\rm and}\quad
	P_{XY}(a,a)\leq P_Y(a)
$$
for every $a\in\mathcal{A}$, 
because of the relationship between a joint probability distribution and the associated marginal distributions.
So, we have
\begin{equation}
	P_{XY}(a,a)\leq \min\{P_X(a),P_Y(a)\}
	\quad
	\forall a\in\mathcal{A}.
	\label{couplinglemma:proof:a:eq000}
\end{equation}
Note that the converse of the inequality above, 
$P_{XY}(a,a)> \min\{P_X(a),P_Y(a)\}$, is never established.
Summing up the both sides of Eq.(\ref{couplinglemma:proof:a:eq000}) with respect to $a$, we have
\begin{eqnarray}
	\Pr\{x=y\}
	&=&
	\sum_{a\in\mathcal{A}}
	P_{XY}(a,a)
	\nonumber\\
	&\leq&
	\sum_{a\in\mathcal{A}}
	\min\{P_X(a),P_Y(a)\}.
\end{eqnarray}
This inequality is arranged to the following form.
\begin{eqnarray}
	\Pr\{x\neq y\}
	&=&
	1-\Pr\{x=y\}
	\nonumber\\
	&\geq&
	1-
	\sum_{a\in\mathcal{A}}
	\min\{P_X(a),P_Y(a)\}
	\nonumber\\
	&=&
	\sum_{a\in\mathcal{A}}
	\Bigl[
		P_X(a)
		-
		\min\{P_X(a),P_Y(a)\}
	\Bigr].
	\nonumber\\
	&&
	\label{couplinglemma:proof:a:eq001}
\end{eqnarray}
Here let us define the following partition of $\mathcal{A}$:
\begin{eqnarray}
	\mathcal{B}
	&=&
	\{b:P_X(b)\geq P_Y(b)\},
	\nonumber\\
	\quad
	\overline{\mathcal{B}}
	&=&
	\{b:P_X(b)< P_Y(b)\}.
	\nonumber
\end{eqnarray}
Then
\begin{eqnarray}
	{\rm Eq.}(\ref{couplinglemma:proof:a:eq001})
	&=&
	\sum_{b\in{\mathcal{B}}}
	\Bigl[
		P_X(b)
		-
		\min\{P_X(b),P_Y(b)\}
	\Bigr]
	\nonumber\\
	&&
	+
	\sum_{b\in\overline{\mathcal{B}}}
	\Bigl[
		P_X(b)
		-
		\min\{P_X(b),P_Y(b)\}
	\Bigr]
	\nonumber\\
	&=&
	\sum_{b\in{\mathcal{B}}}
	\Bigl[
		P_X(b)
		-
		P_Y(b)
	\Bigr]
	\nonumber\\
	&=&
	P_X(\mathcal{B})-P_Y(\mathcal{B})
	\nonumber\\
	&=&
	\max_{\mathcal{S}\subseteq{\mathcal{A}}}
	\Bigl[
		P_X(\mathcal{S})-P_Y(\mathcal{S})
	\Bigr]
	\nonumber\\
	&=&
	v(P_X,P_Y).
	\label{couplinglemme:proof:a:eq008:eq007}
\end{eqnarray}
This completes the proof for (a).

(b) In the preceding part, we observed that the inequality (\ref{couplinglemma:case:a}) follows from Eq.(\ref{couplinglemma:proof:a:eq000}). This implies that the equality in Eq.(\ref{couplinglemma:case:a}) is established when the equality in Eq.(\ref{couplinglemma:proof:a:eq000}) holds. The question is whether there exists a coupling $P_{XY}$ such that
\begin{equation}
	P_{XY}(a,a)=\min\{P_X(a),P_Y(a)\}
	\quad
	\forall a\in\mathcal{A}.
	\label{couplinglemma:proof:b:eq001}
\end{equation}
The existence of such a coupling can be shown by the following constructive method.

\underline{step 1.} Define $P_{XY}(a,a)$ by Eq.(\ref{couplinglemma:proof:b:eq001}). It is clear that $P_{XY}(a,a)\geq 0$ for every $a\in\mathcal{A}$.

\underline{step 2.} 
If $v(P_X,P_Y)=0$, it means that the two random variables, $X$ and $Y$, are equal: $X=Y$. In this case, we let
$
	P_{XY}(a,b)=0
$ for $(a,b)\in\mathcal{A}^2$ such that $a\neq b$.
Clearly, $\Pr\{x\neq y\}=0=v(P_X,P_Y)$, and
$$
	\sum_{y\in\mathcal{A}}P_{XY}(a,y)
	=P_X(a)
	=P_Y(a)
	=
	\sum_{x\in\mathcal{A}}P_{XY}(x,a)
$$
for every $a\in\mathcal{A}$.

Next let us consider the case for $v(P_X,P_Y)\neq 0$.
In this case, the two random variables, $X$ and $Y$, are not equal: $X\neq Y$. This implies $\Pr\{x\neq y\}>0$. Then we define
\begin{equation}
	P_{XY}(a,b)
	=
	\frac{
	R_X(a)R_Y(b)}{\Pr\{x\neq y\}}
	\quad
	\forall (a,b)\in\mathcal{A}^2\mbox{ s.t. }a\neq b,
	\label{couplinglemma:proof:b:eq002}
\end{equation}
where the functions in the numerator are given by
\begin{eqnarray}
	R_X(a)
	&=&
	P_X(a)-P_{XY}(a,a)
	\quad
	\forall a\in\mathcal{A},
	\\
	R_Y(b)
	&=&
	P_Y(b)-P_{XY}(b,b)
	\quad
	\forall b\in\mathcal{A}.
	\label{couplinglemma:proof:b:eq003}
\end{eqnarray}
Since $R_X(a)\geq 0$ and $R_Y(b)\geq 0$, we see that $P_{XY}(a,b)\geq 0$ for every $(a,b)\in\mathcal{A}^2$.
In addition, we observe that
\begin{equation}
	\sum_{a\in\mathcal{A}}R_X(a)
	=
	\sum_{b\in\mathcal{A}}R_Y(b)
	=
	\Pr\{x\neq y\}
	\label{couplinglemma:proof:b:eq004}
\end{equation}
and
\begin{equation}
	R_X(a)R_Y(a)
	=
	R_X(b)R_Y(b)
	=0
	\quad
	\forall a,b\in\mathcal{A}.
	\label{couplinglemma:proof:b:eq005}
\end{equation}
By using Eqs.(\ref{couplinglemma:proof:b:eq004}) and (\ref{couplinglemma:proof:b:eq005}), we can verify that the constructed joint probability distribution has the given distributions $P_X$ and $P_Y$ as the associated marginal distributions. 
For any fixed $a\in\mathcal{A}$, 
\begin{eqnarray}
	& &
	\sum_{b\in\mathcal{A}}
	P_{XY}(a,b)
	\nonumber\\
	&=&
	\sum_{b:b\neq a}
	P_{XY}(a,b)
	+
	\sum_{b:b=a}
	P_{XY}(a,b)
	\nonumber\\
	&=&
	\sum_{b:b\neq a}
	\frac{
	R_X(a)R_Y(b)}{\Pr\{x\neq y\}}
	+
	P_{XY}(a,a)
	\nonumber\\
	&=&
	\frac{R_X(a)}{\Pr\{x\neq y\}}
	\left(
	\sum_{b:b\neq a}
	R_Y(b)
	\right)
	+
	P_{XY}(a,a)
	\nonumber\\
	&=&
	\frac{R_X(a)}{\Pr\{x\neq y\}}
	\left(
	\sum_{b\in\mathcal{A}}
	R_Y(b)
	-R_Y(a)
	\right)
	+
	P_{XY}(a,a)
	\nonumber\\
	&=&
	\frac{R_X(a)}{\Pr\{x\neq y\}}
	\left(
	\Pr\{x\neq y\}
	-R_Y(a)
	\right)
	+
	P_{XY}(a,a)
	\nonumber\\
	&=&
	R_X(a)
	-\frac{R_X(a)R_Y(a)}{\Pr\{x\neq y\}}
	+
	P_{XY}(a,a)
	\nonumber\\
	\nonumber\\
	&=&
	R_X(a)
	+
	P_{XY}(a,a)
	\nonumber\\
	&=&
	P_X(a).
	\nonumber
\end{eqnarray}
With the same manner, we also have
$$
	\sum_{a\in\mathcal{A}}
	P_{XY}(a,b)
	=
	P_Y(b)
	\quad
	\forall b\in\mathcal{A}.
$$
Thus, it was shown that the joint probability distribution $P_{XY}$ constructed from Eqs.(\ref{couplinglemma:proof:b:eq001}) and (\ref{couplinglemma:proof:b:eq002}) has the marginal distributions $P_X$ and $P_Y$. Since Eq.(\ref{couplinglemma:proof:b:eq001}) holds, we have
$v(P_X,P_Y)=\Pr\{x\neq y\}$ with this constructed distribution $P_{XY}$.
\hfill$\blacksquare$

\vspace{1em}
A concrete example of couplings of $P_X$ and $P_Y$ is shown in the appendix {\it A}. This illustrates the coupling lemma ({\it Theorem 3}) in a case of $X\neq Y$.

It should be emphasized that the coupling lemma consists of two parts: (a) ``for every" part and (b) ``there exists" part. Lemma 1 of the literature \cite{RK05}, Proposition 2.1.1 of the literature \cite{Ren08}, and Theorem A.6 of the literature \cite{PR14}, these are essentially identical to the ``there exists" part of the coupling lemma. 
On the other hand, an example of independent joint probability distribution, which was treated in the literatures \cite{Y09,Y10}, 
can be explained by the ``for every" part of the coupling lemma. For the later discussion, we treat the independent joint distribution case here again.

Suppose that $X$ and $Y$ are independent. 
Then the joint probability distribution $P_{XY}$ is given by 
\begin{equation*}
	P_{XY}(a,b)=P_X(a)P_Y(b)
	\quad
	\forall (a,b)\in\mathcal{A}^2.
\end{equation*}
Since $0\leq P_X(a)\leq 1$ and $0\leq P_Y(b)\leq 1$, 
we have
\begin{equation*}
	P_X(a)P_Y(a)\leq \min\{P_X(a),P_Y(a)\}
	\quad
	\forall a\in\mathcal{A}.
\end{equation*}
If some $a^\prime\in\mathcal{A}$ satisfies the conditions
$0<P_X(a^\prime)<1$ and 
$0<P_Y(a^\prime)<1$, 
then it is reduced to
\begin{equation}
	P_X(a^\prime)P_Y(a^\prime)< \min\{P_X(a^\prime),P_Y(a^\prime)\}.
\end{equation}
Taking this fact into account, we have
\begin{equation*}
	1-\sum_{a\in\mathcal{A}}\min\{P_X(a),P_Y(a)\}
	<
	1-\sum_{a\in\mathcal{A}}P_X(a)P_Y(a),
\end{equation*}
when at least one $a^\prime\in\mathcal{A}$ satisfies the conditions
$0<P_X(a^\prime)<1$ and 
$0<P_Y(a^\prime)<1$. 
With the help of the calculation of Eq.(\ref{couplinglemme:proof:a:eq008:eq007}), the above inequality is summarized to the following statement:

\vspace{1em}
\noindent
{\it Corollary to Theorem 3}:
Suppose that $X$ and $Y$ are independent, and at least one $a^\prime\in\mathcal{A}$ satisfies the conditions
$0<P_X(a^\prime)<1$ and 
$0<P_Y(a^\prime)<1$. Then, 
\begin{equation}
	v(P_X,P_Y)<\Pr\{x\neq y\}.
\end{equation}
\hfill$\blacksquare$

\vspace{1em}
The next example was used in the literatures \cite{Y09,Y10} to explain the incorrectness of the failure probability interpretation.

\vspace{1em}
\noindent
{\it Example 4} (\cite{Y09,Y10}):
Let $\mathcal{A}=\{1,2,\dots,N\}$, and 
$P_X(x)=1/N\ \forall x\in{\mathcal{A}}$ and 
$P_Y(y)=1/N\ \forall y\in{\mathcal{A}}$.
Suppose that $X$ and $Y$ are independent. Then $P_{XY}(x,y)=P_X(x)P_Y(y)=1/N^2$ for every $(x,y)\in\mathcal{A}^2$. In this case, we have
$$
	 v(P_X,P_Y)=0 <
	 1-\frac{1}{N} = \Pr\{x\neq y\}
$$
for $N\geq 2$.
\hfill$\blacksquare$

\vspace{1em}
Up to this point, we have considered the case of one-dim random variables.
It is easy to extend the case of $v(P_X,P_Y)$ to that of $v(P_{X_1X_2},P_{Y_1Y_2})$.

\vspace{1em}
\noindent
{\it Theorem 5}:
Let $(X_1,X_2)$ and $(Y_1,Y_2)$ are two-dim random variables that take on values from the same finite alphabet $\mathcal{A}^2$, where $\mathcal{A}=\{a_1,a_2,\dots,a_N\}$. 
Suppose that $P_{X_1X_2}$ and $P_{Y_1Y_2}$ are given.
\begin{itemize}
	\item[(a)] 
	{\it For every} coupling $P_{X_1X_2Y_1Y_2}$ of $P_{X_1X_2}$ and $P_{Y_1Y_2}$, 
	\begin{equation}
		v(P_{X_1X_2},P_{Y_1Y_2})
		\leq
		\Pr\{(x_1,x_2)\neq (y_1,y_2)\}
		\label{couplinglemma:case:a:extension}
	\end{equation}
	where
	\begin{eqnarray}
	&&
	\!\!\!\!
	\!\!\!\!
	\!\!\!\!
	\!\!\!\!
	\!\!\!\!
	\!\!\!\!
	v(P_{X_1X_2},P_{Y_1Y_2})
	\nonumber\\
	&=&
	\frac{1}{2}\sum_{a\in\mathcal{A}}\sum_{b\in\mathcal{A}}
	\Bigl|
	P_{X_1X_2}(a,b)-P_{Y_1Y_2}(a,b)
	\Bigr|.
	\nonumber
\end{eqnarray}
	\item[(b)] 
	{\it There exists} a coupling $P_{X_1X_2Y_1Y_2}$ such that
	\begin{equation}
		v(P_{X_1X_2},P_{Y_1Y_2})
		=
		\Pr\{(x_1,x_2)\neq (y_1,y_2)\}.
		\label{couplinglemma:case:b:extension}
	\end{equation}
\end{itemize}
\hfill$\square$

\noindent
{\it Proof}:
This is due to the inequality
\begin{eqnarray}
	&&
	\!\!\!\!
	\!\!\!\!
	\!\!\!\!
	\!\!\!\!
	\!\!\!\!
	\!\!\!\!
	P_{X_1X_2Y_1Y_2}(a,b,a,b)
	\nonumber\\
	&\leq&
	\min\{
	P_{X_1X_2}(a,b),
	P_{Y_1Y_2}(a,b)
	\}
	\label{proof:eq022:coupling:extended}
\end{eqnarray}
for every $(a,b)\in\mathcal{A}^2$.
\hfill$\blacksquare$

\vspace{1em}
A concrete example of couplings of $P_{X_1X_2}$ and $P_{Y_1Y_2}$ is shown in the appendix {\it B}. This illustrates the coupling lemma for two-dim random variables ({\it Theorem 5}) in a case of $(X_1,X_2)\neq (Y_1,Y_2)$.

The reader might recall a description on the variational distance for two-dim random variables $(\tilde{X},X)$ and $(\tilde{X},Y)$ in the textbook \cite{NC10} of Nielsen and Chuang (p.402). From the point of view of the coupling lemma, it can be understood as a special case that has the condition $X_1=X_2=Y_1$.

\vspace{1em}
\noindent
{\it Corollary to Theorem 5}:
When the condition $X_1=X_2=Y_1$ is imposed, then 
\begin{equation}
	v(P_{X_1X_2},P_{Y_1Y_2})
	=
	\Pr\{x_2\neq y_2\}.
\end{equation}
\hfill$\square$

\noindent
{\it Proof}:
From the condition $X_1=X_2=Y_1$, we have
$P_{X_1X_2Y_1Y_2}(x_1,x_2,y_1,b)=0$
if $x_2\neq x_1$ or $y_1\neq x_1$.
Hence, for every $a,b\in\mathcal{A}$, 
\begin{eqnarray}
	P_{X_1Y_2}(a,b)
	&=&
	\sum_{x_2\in\mathcal{A}}
	\sum_{y_1\in\mathcal{A}}
	P_{X_1X_2Y_1Y_2}(a,x_2,y_1,b)
	\nonumber\\
	&=&
	P_{X_1X_2Y_1Y_2}(a,a,a,b);
	\\
	P_{X_2Y_2}(a,b)
	&=&
	\sum_{x_1\in\mathcal{A}}
	\sum_{y_1\in\mathcal{A}}
	P_{X_1X_2Y_1Y_2}(x_1,a,y_1,b)
	\nonumber\\
	&=&
	P_{X_1X_2Y_1Y_2}(a,a,a,b);
	\\
	P_{Y_1Y_2}(a,b)
	&=&
	\sum_{x_1\in\mathcal{A}}
	\sum_{x_2\in\mathcal{A}}
	P_{X_1X_2Y_1Y_2}(x_1,x_2,a,b)
	\nonumber\\
	&=&
	P_{X_1X_2Y_1Y_2}(a,a,a,b).
	\label{eq015}
\end{eqnarray}
If $a=b$, then
\begin{eqnarray}
	P_{X_1X_2}(a,b)
	&=&
	P_{X_1X_2}(a,a)
	\nonumber\\
	&=&
	\sum_{y_1\in\mathcal{A}}
	\sum_{y_2\in\mathcal{A}}
	P_{X_1X_2Y_1Y_2}(a,a,y_1,y_2)
	\nonumber\\
	&=&
	\sum_{y_2\in\mathcal{A}}
	P_{X_1X_2Y_1Y_2}(a,a,a,y_2)
	\nonumber\\
	&\geq&
	P_{X_1X_2Y_1Y_2}(a,a,a,y_2^\prime)
	\quad
	\forall
	y_2^\prime\in\mathcal{A}.
	\nonumber
\end{eqnarray}
This implies
$$
	P_{X_1X_2}(a,b)
	\geq
	P_{X_1X_2Y_1Y_2}(a,a,a,b)
	\quad
	\mbox{if $a=b$}.
$$
Substituting Eq.(\ref{eq015}) into this, we have
$$
	P_{X_1X_2}(a,a)\geq P_{X_1X_2Y_1Y_2}(a,a,a,a)=P_{Y_1Y_2}(a,a).
$$
Therefore, 
\begin{eqnarray}
	&&
	\!\!\!\!
	\!\!\!\!
	\!\!\!\!
	\!\!\!\!
	\!\!\!\!
	\!\!\!\!
	P_{X_1X_2Y_1Y_2}(a,a,a,a)
	\nonumber\\
	&=&
	P_{Y_1Y_2}(a,a)
	\nonumber\\
	&=&
	\min
	\{
	P_{X_1X_2}(a,a),
	P_{Y_1Y_2}(a,a)
	\}.
	\label{eq019}
\end{eqnarray}

Next we assume that $a\neq b$. It is obvious that
$P_{X_1X_2Y_1Y_2}(a,b,a,b)=0=P_{X_1X_2}(a,b)$.
By using Eq.(\ref{eq015}),
\begin{eqnarray}
	P_{Y_1Y_2}(a,b)
	&=&
	P_{X_1X_2Y_1Y_2}(a,a,a,b)
	\nonumber\\
	&\geq&
	0
	\nonumber\\
	&=&
	P_{X_1X_2}(a,b)
	\nonumber\\
	&=&
	P_{X_1X_2Y_1Y_2}(a,b,a,b).
	\label{eq020}
\end{eqnarray}
This yields
\begin{eqnarray}
	&&
	\!\!\!\!
	\!\!\!\!
	\!\!\!\!
	\!\!\!\!
	\!\!\!\!
	\!\!\!\!
	P_{X_1X_2Y_1Y_2}(a,b,a,b)
	\nonumber\\
	&=&
	P_{X_1X_2}(a,b)
	\nonumber\\
	&=&
	\min
	\{
	P_{X_1X_2}(a,b),
	P_{Y_1Y_2}(a,b)
	\}.
	\label{eq021}
\end{eqnarray}
Summarizing Eqs.(\ref{eq019}) and (\ref{eq021}), 
\begin{eqnarray}
	&&
	\!\!\!\!
	\!\!\!\!
	\!\!\!\!
	\!\!\!\!
	\!\!\!\!
	\!\!\!\!
	P_{X_1X_2Y_1Y_2}(a,b,a,b)
	\nonumber\\
	&=&
	\min
	\{
	P_{X_1X_2}(a,b),
	P_{Y_1Y_2}(a,b)
	\}
	\label{eq022}
\end{eqnarray}
for every $(a,b)\in\mathcal{A}^2$.
From the comparison between this and Eq.(\ref{proof:eq022:coupling:extended}), we find that
\begin{equation*}
	v(P_{X_1X_2},P_{Y_1Y_2})
	=
	\Pr\{(x_1,x_2)\neq (y_1,y_2)\}.
\end{equation*}
Moreover, since $X_1=X_2=Y_1$, the probability
$
	\Pr\{(x_1,x_2)\neq (y_1,y_2)\}
$ is reduced to 
$
	\Pr\{x_2\neq y_2\}
$.
Thus, we proved that $X_1=X_2=Y_1$ yields the equality of
$v(P_{X_1X_2},P_{Y_1Y_2})$ and $\Pr\{x_2\neq y_2\}$.
\hfill$\blacksquare$

\vspace{1em}
Finally, let us summarize the relationship between the variational distance and the probability of $x\neq y$.
\begin{enumerate}
	\item The variational distance $v(P_X,P_Y)$ does not always mean the probability $\Pr\{x\neq y\}$. $v(P_{X_1X_2},P_{Y_1Y_2})$ too. Basically, the variational distance $v(P_X,P_Y)$ is a {\it lower} bound of $\Pr\{x\neq y\}$.
	\item Suppose that the two random variables $X$ and $Y$ are independent, and some $a^\prime\in\mathcal{A}$ satisfies the conditions
$0<P_X(a^\prime)<1$ and 
$0<P_Y(a^\prime)<1$. 
In this case, we have $v(P_X,P_Y)<\Pr\{x\neq y\}$.
	\item Conversely, a maximal coupling of $P_X$ and $P_Y$ that yields $v(P_X,P_Y)=\Pr\{x\neq y\}$ demands correlation between the outcomes of $X$ and $Y$.
\end{enumerate}
\section{On The Failure Probability Interpretation For $\varepsilon$}
Let us return to the main course of this paper. 
In the security analysis of QKD under the trace distance criterion, as mentioned in the section I, 
``$\varepsilon$-secure" is defined by
\begin{equation*}
	d=\frac{1}{2}
	\|\hat{\boldsymbol{\rho}}_{KE}-\hat{\rho}_U\otimes\hat{\rho}_{E}\|
	\leq \varepsilon.
\end{equation*}
The problem that Yuen and Hirota have discussed and we discuss here is whether or not the parameter $\varepsilon$ has a meaning of ``the failure probability".

In the literatures \cite{RK05,KRBM07,Ren08,SR08,MR09,SBCDLP09,TLGR12,PR14}, it is claimed that the operational meaning of $\varepsilon$ is the (maximal) failure probability of the QKD protocol. To give such an interpretation, they first make the inequality
\begin{equation}
	v(P_K, P_U)\leq d\leq \varepsilon,
	\label{discussion:renner}
\end{equation}
where $P_K$ is a probability distribution of the {\it real} key and $P_U$ is that of the {\it ideal} key (uniform key). After that, they apply the statement (b) of the coupling lemma to give the failure probability interpretation. 
If we dare to ignore the physical restrictions and focus only on the mathematical possibilities of the coupling lemma, a coupling $P_{KU}$ of $P_K$ and $P_U$ that yields $v(P_K, P_U)=\Pr\{k\neq u\}$ exists. However, we cannot drop physical restrictions. 

The distribution $P_K$ describes the probabilistic behavior of measurement outcomes that are obtained through an appropriate POVM for the system of $\hat{\boldsymbol{\rho}}_{KE}$. In addition, the distribution $P_U$ is obtained by the same POVM. If one supposes that the measurement outcomes of the {\it real} system and the {\it ideal} system are characterized by a maximal coupling, it means that measurement outcomes from the two distinct systems are correlated. In other words, it demands that measurement outcomes in the {\it real} system depends on that in the {\it ideal} system. 
There is no clear reason why one must choose such a coupling and why correlation between the {\it real} keys and the {\it ideal} keys is allowed. 

If the equality of the variational distance $v(P_K,P_U)$ and the probability $\Pr\{k\neq u\}$ were established for {\it every} coupling of $P_K$ and $P_U$, the failure probability interpretation might be justified. But, we have already seen that the equality does not always hold.

Further, we assume here that the measurement outcomes from the {\it real} and {\it ideal} systems are independent. It would be a natural situation, and is physically acceptable.  As shown by Shannon \cite{Shannon49} (p.681), all keys have to be equally likely to make one-time pad secure. This is reflected to the settings of the {\it ideal} probability distribution $P_U$. Indeed, $P_U$ is given to be a uniform distribution.  So, $P_U$ satisfies the condition $0<P_U(u)<1$. %
On the other hand, if some sequence $k^\prime$ generated by the QKD protocol possesses probability $P_K(k^\prime)=0$ or $P_K(k^\prime)=1$, it is clear that the generated sequence does not work as the secret key for the embedded one-time pad. Therefore, we can assume that every sequence $k$ satisfies the condition $0<P_K(k)<1$ without loss of generality. But, as shown in {\it Corollary to Theorem 3}, the strict inequality 
\begin{equation}
	v(P_K,P_U)<\Pr\{k\neq u\}
	\label{discussion:independent}
\end{equation}
is established in this case. 
The juxtaposition of Eqs.(\ref{discussion:renner}) and (\ref{discussion:independent}) tells nothing about the relationship between $\varepsilon$ and $\Pr\{k\neq u\}$.

Consequently, we have the following facts:
\begin{enumerate}
	\item If a maximal coupling of $P_K$ and $P_U$ is employed, correlation between the {\it real} and {\it ideal} keys is needed. But, there is no cause one must choose a maximal coupling.
	\item $v(P_K,P_U)\leq \Pr\{k\neq u\}$ holds for every coupling of $P_K$ and $P_U$. That is, $v(P_K,P_U)$ is a lower bound of all the possible probabilities $\Pr\{k\neq u\}$.
	\item If the {\it real} and {\it ideal} keys are statistically independent, $v(P_K,P_U)< \Pr\{k\neq u\}$. Clearly, $v(P_K,P_U)$ does not mean $\Pr\{k\neq u\}$.
\end{enumerate}

Let us recall {\it Example 4}. 
In the viewpoint of the application of the coupling lemma, we can understand that
this example illustrates all the facts listed above compactly; it immediately shows the fact 3), and by considering the ``not independent" cases, one would arrive at the facts 1) and 2).
Thus, the use of the coupling lemma justifies the Yuen's claim for the failure probability interpretation problem that are repeatedly stated in the literatures \cite{Y09,Y10} and its subsequent papers. 
In summary, we can say that $\varepsilon$ does not mean probability at all, in particular, it does not mean ``the failure probability".


\section{Conclusions} 
This paper is concerned with the failure probability interpretation problem that has been presented by Yuen and Hirota. To discuss the problem, we used the coupling lemma. First, we observed that the variational distance does not always mean probability due to the coupling lemma. In particular, the analysis of the independent distribution case showed that the variational distance of such a case is not a probability. Applying the coupling lemma to the discussions on the failure probability interpretation problem, we could justify the Yuen's claim stated in the literatures \cite{Y09,Y10} and its subsequent papers. As a result, we can conclude that the failure probability interpretation is not adequate in the security analysis of quantum key distribution.

\section{Acknowledgment}
The author would like to thank Osamu Hirota for fruitful discussions and suggestions.


\vspace{3em}
\appendix

\vspace{3em}
\subsection{Coupling of $P_X$ and $P_Y$}
In this section we treat a case of $X\neq Y$ to illustrate the mathematical notion of the coupling of two probability distributions. 

Suppose that the distributions of $X$ and $Y$ are respectively given as
\begin{eqnarray}
	P_X
	&=&
	(0.10000,0.20000,0.30000,0.40000),
	\nonumber
	\\
	P_Y
	&=&
	(0.25000,0.25000,0.25000,0.25000).
	\nonumber
\end{eqnarray}
Then we have $v(P_X,P_Y)=0.20000$.

\vspace{2em}
\subsubsection{Case a ($X$ and $Y$ are independent)} 
If the random variables $X$ and $Y$ are independent, then the joint probability distribution is given as follows.
\begin{equation}
	P_{XY}^{(a)}
	=
	\left(
	\begin{array}{cccc}
		0.02500 & 0.02500 & 0.02500 & 0.02500 \\
		0.05000 & 0.05000 & 0.05000 & 0.05000 \\
		0.07500 & 0.07500 & 0.07500 & 0.07500 \\
		0.10000 & 0.10000 & 0.10000 & 0.10000
	\end{array}
	\right).
	\nonumber
\end{equation}
With this joint probability distribution, we have
$$
	v(P_X,P_Y)
	=
	0.20000
	<
	0.75000
	=
	\Pr\{x\neq y\}.
$$
This is also an example for {\it Corollary to Theorem 3}. Thus the variational distance $v(P_X,P_Y)$ behaves as a lower bound of the probability $\Pr\{x\neq y\}$ when $X$ and $Y$ are independent.

\vspace{2em}
\subsubsection{Case b (maximal coupling)}
Using the construction procedure shown in Section II, 
a maximal coupling of $P_X$ and $P_Y$ is given by
\begin{equation}
	P_{XY}^{(b)}
	=
	\left(
	\begin{array}{cccc}
		0.10000 & 0.00000 & 0.00000 & 0.00000 \\
		0.00000 & 0.20000 & 0.00000 & 0.00000 \\
		0.03750 & 0.01250 & 0.25000 & 0.00000 \\
		0.11250 & 0.03750 & 0.00000 & 0.25000
	\end{array}
	\right).
	\nonumber
\end{equation}
In this case, we have
$$
	v(P_X,P_Y)
	=
	0.20000
	=
	\Pr\{x\neq y\},
$$
which is an example of {\it Theorem 3}(b).
Observe that this coupling demands a strong correlation between $X$ and $Y$.
Indeed the following events never happen if $X$ and $Y$ obey this coupling:
$$
	\begin{array}{rl}
	\mbox{ event} & (X=1)\wedge (Y=2),\\
	\mbox{ event} & (X=1)\wedge (Y=3),\\
	\mbox{ event} & (X=1)\wedge (Y=4),\\
	\mbox{ event} & (X=2)\wedge (Y=1),\\
	\mbox{ event} & (X=2)\wedge (Y=3),\\
	\mbox{ event} & (X=2)\wedge (Y=4),\\
	\mbox{ event} & (X=3)\wedge (Y=4),\mbox{ and}\\
	\mbox{ event} & (X=4)\wedge (Y=4).
	\end{array}
$$
Thus the half of the possible events never occur.

\vspace{2em}
\subsubsection{Case c (not independent, not maximal)}
When two probability distributions are given, there are infinitely many couplings in general. {\it Theorem 3}(a) covers all of the possible couplings, so that no coupling breaks the inequality (\ref{couplinglemma:case:a}). 
To see this, let us consider the following joint probability distribution of $P_X$ and $P_Y$.
\begin{equation}
	P_{XY}^{(c)}
	=
	\left(
	\begin{array}{cccc}
	0.06250 & 0.01250 & 0.01250 & 0.01250 \\
	0.02500 & 0.12500 & 0.02500 & 0.02500 \\
	0.05625 & 0.04375 & 0.16250 & 0.03750 \\
	0.10625 & 0.06875 & 0.05000 & 0.17500
	\end{array}
	\right).
	\nonumber
\end{equation}
This is neither the case of independent random variables nor the case of maximal couplings. Even in this case, of course, the inequality holds.
$$
	v(P_X,P_Y)
	=
	0.20000
	<
	0.47500
	=
	\Pr\{x\neq y\}.
$$
This is a typical behavior of the variational distance.

\vspace{2em}
\subsection{Coupling of $P_{X_1X_2}$ and $P_{Y_1Y_2}$}
This section gives a concrete example of couplings for two two-dim probability distributions.

Suppose that the distributions of $(X_1,X_2)$ and $(Y_1,Y_2)$ are respectively given as
\begin{equation}
	P_{X_1X_2}
	=
	\left(
\begin{array}{ccc}
1/3 & 0 & 0 \\
 0 & 1/3 & 0 \\
 0 & 0 & 1/3
\end{array}
\right),
\end{equation}
and
\begin{equation}
	P_{Y_1Y_2}
	=
	\left(
\begin{array}{ccc}
 1/9 & 2/9 & 0 \\
 1/9 & 1/9 & 1/9 \\
 0 & 1/9 & 2/9
\end{array}
\right).
\end{equation}
For these distributions, $v(P_{X_1X_2},P_{Y_1Y_2})=5/9$.

\vspace{2em}
\subsubsection{Case a (maximal coupling)}
The joint probability $P^{(a)}_{X_1X_2Y_1Y_2}$ in TABLE I (of the next page) is constructed according to the procedure described in Section II. This provides a maximal coupling of $P_{X_1X_2}$ and $P_{Y_1Y_2}$. In this case we have
$$
	v(P_{X_1X_2},P_{Y_1Y_2})
	=
	\frac{5}{9}
	=
	\Pr\{(x_1,x_2)\neq(y_1,y_2)\}.
$$

\vspace{2em}
\subsubsection{Case b ($X_1=X_2=Y_1$ is imposed)}
The joint distribution $P^{(b)}_{X_1X_2Y_1Y_2}$ in TABLE II (of the next page) is designed for satisfying the condition $X_1=X_2=Y_1$. For this joint distribution, we have
$$
	v(P_{X_1X_2},P_{Y_1Y_2})
	=
	\frac{5}{9}
	=
	\Pr\{(x_1,x_2)\neq(y_1,y_2)\}.
$$
Observe that
$$
	\Pr\{(x_1,x_2)\neq(y_1,y_2)\}=\Pr\{x_2 \neq y_2\}=\frac{5}{9}.
$$
Therefore we have
$$
	v(P_{X_1X_2},P_{Y_1Y_2})
	=
	\frac{5}{9}
	=
	\Pr\{x_2 \neq y_2\}.
$$
This can also be understood as an example for the case of $X=\tilde{X}$ in the textbook \cite{NC10} of Nielsen and Chuang.

\vspace{2em}
\subsubsection{Case c ($(X_1,X_2)$ and $(Y_1,Y_2)$ are independent)}
Let us consider the case that $(X_1,X_2)$ and $(Y_1,Y_2)$ are independent.
The joint distribution $P^{(c)}_{X_1X_2Y_1Y_2}$ for this case is shown in TABLE III (of the next page). As expected from {\it Corollary to Theorem 3}, we have the following strict inequality.
$$
	v(P_{X_1X_2},P_{Y_1Y_2})
	=
	\frac{5}{9}
	<
	\frac{23}{27}
	=
	\Pr\{(x_1,x_2)\neq(y_1,y_2)\}.
$$
Note that this example also implies that if the condition $X_1=X_2$ is only imposed, that is, the condition $X_1=Y_1$ is removed from the condition $X_1=X_2=Y_1$, then the equality of
$v(P_{X_1X_2},P_{Y_1Y_2})$ and $\Pr\{x_2\neq y_2\}$ is not guaranteed.

\begin{table*}[p]
	\caption{Maximal coupling $P^{(a)}_{X_1X_2Y_1Y_2}(x_1,x_2,y_1,y_2)$ constructed by the procedure described in Section II.}
	\label{tab001}
	\centerline{
		\begin{tabular}{|c|c||c|c|c|c|c|c|c|c|c|}
			\cline{3-11}
			\multicolumn{2}{c}{\hspace{1em}} &
			\multicolumn{3}{|c|}{$Y_1=1$}&
			\multicolumn{3}{|c|}{$Y_1=2$}&
			\multicolumn{3}{|c|}{$Y_1=3$}\\
			\cline{3-11}
			\multicolumn{2}{c|}{\hspace{1em}} &
			$Y_2=1$ & $Y_2=2$ & $Y_2=3$ &
			$Y_2=1$ & $Y_2=2$ & $Y_2=3$ &
			$Y_2=1$ & $Y_2=2$ & $Y_2=3$ \\
			\hline\hline
			& $X_2=1$ &
			$1/9$ & $4/45$ & $0$ &
			$2/45$ & $0$ & $2/45$ &
			$0$ & $2/45$ & $0$ \\
			\cline{2-11}
			$X_1=1$ & $X_2=2$ &
			$0$ & $0$ & $0$ &
			$0$ & $0$ & $0$ &
			$0$ & $0$ & $0$ \\
			\cline{2-11}
			& $X_2=3$ &
			$0$ & $0$ & $0$ &
			$0$ & $0$ & $0$ &
			$0$ & $0$ & $0$ \\
			\hline
			& $X_2=1$ &
			$0$ & $0$ & $0$ &
			$0$ & $0$ & $0$ &
			$0$ & $0$ & $0$ \\
			\cline{2-11}
			$X_1=2$ & $X_2=2$ &
			$0$ & $4/45$ & $0$ &
			$2/45$ & $1/9$ & $2/45$ &
			$0$ & $2/45$ & $0$ \\
			\cline{2-11}
			& $X_2=3$ &
			$0$ & $0$ & $0$ &
			$0$ & $0$ & $0$ &
			$0$ & $0$ & $0$ \\
			\hline
			& $X_2=1$ &
			$0$ & $0$ & $0$ &
			$0$ & $0$ & $0$ &
			$0$ & $0$ & $0$ \\
			\cline{2-11}
			$X_1=2$ & $X_2=2$ &
			$0$ & $0$ & $0$ &
			$0$ & $0$ & $0$ &
			$0$ & $0$ & $0$ \\
			\cline{2-11}
			& $X_2=3$ &
			$0$ & $2/45$ & $0$ &
			$1/45$ & $0$ & $1/45$ &
			$0$ & $1/45$ & $2/9$ \\
			\hline
		\end{tabular}
	}
	\vspace{8em}
	\caption{Coupling $P^{(b)}_{X_1X_2Y_1Y_2}(x_1,x_2,y_1,y_2)$ under the condition $X_1=X_2=Y_1$}
	\label{tab002}
	\centerline{
		\begin{tabular}{|c|c||c|c|c|c|c|c|c|c|c|}
			\cline{3-11}
			\multicolumn{2}{c}{\hspace{1em}} &
			\multicolumn{3}{|c|}{$Y_1=1$}&
			\multicolumn{3}{|c|}{$Y_1=2$}&
			\multicolumn{3}{|c|}{$Y_1=3$}\\
			\cline{3-11}
			\multicolumn{2}{c|}{\hspace{1em}} &
			$Y_2=1$ & $Y_2=2$ & $Y_2=3$ &
			$Y_2=1$ & $Y_2=2$ & $Y_2=3$ &
			$Y_2=1$ & $Y_2=2$ & $Y_2=3$ \\
			\hline
			\hline
			& $X_2=1$ &
			$1/9$ & $2/9$ & $0$ &
			$0$ & $0$ & $0$ &
			$0$ & $0$ & $0$ \\
			\cline{2-11}
			$X_1=1$ & $X_2=2$ &
			$0$ & $0$ & $0$ &
			$0$ & $0$ & $0$ &
			$0$ & $0$ & $0$ \\
			\cline{2-11}
			& $X_2=3$ &
			$0$ & $0$ & $0$ &
			$0$ & $0$ & $0$ &
			$0$ & $0$ & $0$ \\
			\hline
			& $X_2=1$ &
			$0$ & $0$ & $0$ &
			$0$ & $0$ & $0$ &
			$0$ & $0$ & $0$ \\
			\cline{2-11}
			$X_1=2$ & $X_2=2$ &
			$0$ & $0$ & $0$ &
			$1/9$ & $1/9$ & $1/9$ &
			$0$ & $0$ & $0$ \\
			\cline{2-11}
			& $X_2=3$ &
			$0$ & $0$ & $0$ &
			$0$ & $0$ & $0$ &
			$0$ & $0$ & $0$ \\
			\hline
			& $X_2=1$ &
			$0$ & $0$ & $0$ &
			$0$ & $0$ & $0$ &
			$0$ & $0$ & $0$ \\
			\cline{2-11}
			$X_1=2$ & $X_2=2$ &
			$0$ & $0$ & $0$ &
			$0$ & $0$ & $0$ &
			$0$ & $0$ & $0$ \\
			\cline{2-11}
			& $X_2=3$ &
			$0$ & $0$ & $0$ &
			$0$ & $0$ & $0$ &
			$0$ & $1/9$ & $2/9$ \\
			\hline
		\end{tabular}
	}
	\vspace{8em}
	\caption{Coupling $P^{(c)}_{X_1X_2Y_1Y_2}(x_1,x_2,y_1,y_2)$ under the conditions $(X_1,X_2)$ and $(Y_1,Y_2)$ are independent.}
	\label{tab003}
	\centerline{
		\begin{tabular}{|c|c||c|c|c|c|c|c|c|c|c|}
			\cline{3-11}
			\multicolumn{2}{c}{\hspace{1em}} &
			\multicolumn{3}{|c|}{$Y_1=1$}&
			\multicolumn{3}{|c|}{$Y_1=2$}&
			\multicolumn{3}{|c|}{$Y_1=3$}\\
			\cline{3-11}
			\multicolumn{2}{c|}{\hspace{1em}} &
			$Y_2=1$ & $Y_2=2$ & $Y_2=3$ &
			$Y_2=1$ & $Y_2=2$ & $Y_2=3$ &
			$Y_2=1$ & $Y_2=2$ & $Y_2=3$ \\
			\hline\hline
			& $X_2=1$ &
			$1/27$ & $2/27$ & $0$ &
			$1/27$ & $1/27$ & $1/27$ &
			$0$ & $1/27$ & $2/27$ \\
			\cline{2-11}
			$X_1=1$ & $X_2=2$ &
			$0$ & $0$ & $0$ &
			$0$ & $0$ & $0$ &
			$0$ & $0$ & $0$ \\
			\cline{2-11}
			& $X_2=3$ &
			$0$ & $0$ & $0$ &
			$0$ & $0$ & $0$ &
			$0$ & $0$ & $0$ \\
			\hline
			& $X_2=1$ &
			$0$ & $0$ & $0$ &
			$0$ & $0$ & $0$ &
			$0$ & $0$ & $0$ \\
			\cline{2-11}
			$X_1=2$ & $X_2=2$ &
			$1/27$ & $2/27$ & $0$ &
			$1/27$ & $1/27$ & $1/27$ &
			$0$ & $1/27$ & $2/27$ \\
			\cline{2-11}
			& $X_2=3$ &
			$0$ & $0$ & $0$ &
			$0$ & $0$ & $0$ &
			$0$ & $0$ & $0$ \\
			\hline
			& $X_2=1$ &
			$0$ & $0$ & $0$ &
			$0$ & $0$ & $0$ &
			$0$ & $0$ & $0$ \\
			\cline{2-11}
			$X_1=2$ & $X_2=2$ &
			$0$ & $0$ & $0$ &
			$0$ & $0$ & $0$ &
			$0$ & $0$ & $0$ \\
			\cline{2-11}
			& $X_2=3$ &
			$1/27$ & $2/27$ & $0$ &
			$1/27$ & $1/27$ & $1/27$ &
			$0$ & $1/27$ & $2/27$ \\
			\hline
		\end{tabular}
	}
	\vspace{2em}
	
\end{table*}


\vspace{2em}
\begin{thebibliography}{99}
\bibitem{A83}
	 D.~Aldous, 
	``Random walks on finite groups and rapidly mixing Markov chains,"
	S\'eminaire de Probabilit\'es XVII 1981/82, 
	Lecture Notes in Mathematics, vol.986, pp.243-297, 1983.
\bibitem{RK05}
	R.~Renner and R.~K\"onig, 
	``Universally composable privacy amplification against quantum adversaries," 
	TCC 2005, LNCS 3378, pp.407-425, 2005.
\bibitem{KRBM07}
	R.~K\"onig, R.~Renner, A.~Briska, and U.~Maurer,
	``Small accessible quantum information does not imply security,"
	Phys. Rev. Lett., vol.98, no.14, 140502, 2007.
\bibitem{Ren08}
	R.~Renner, 
	``Security of quantum key distribution," 
	Int. J. Quant. Inform., vol.6, no.1, pp.1-127, 2008.
\bibitem{SR08}
	V.~Scarani, and R.~Renner, 
	``Quantum cryptography with finite resources: Unconditional security bound for discrete-variable protocols with one-way postprocessing,"
	Phys. Rev. Lett., vol.100, no.20, 200501, 2008.
\bibitem{MR09}
	J.~M\"uller-Quade, and R.~Renner, 
	``Composability in quantum cryptography," 
	New J. Phys., vol.11, 085006, 2009.
\bibitem{SBCDLP09}
	V.~Scarani, 
	H.~Bechmann-Pasquinucci,
	N.~J.~Cerf,
	M.~Du\v{s}ek,
	N.~L\"utkenhaus,
	and
	M.~Peev,
	``The security of practical quantum key distribution,"
	Rev. Mod. Phys., vol.81, no.3, pp.1301-1350, 2009.
\bibitem{TLGR12}
	M.~Tomamichel, C.~C.~W.~Lim, N.~Gisin, R.~Renner,
	``Tight finite-key analysis for quantum cryptography,"
	Nature Commun, vol.3, 634, 2012.
\bibitem{PR14}
	C.~Portmann, and R.~Renner, 
	``Cryptographic security of quantum key distribution,"
	quant-ph arXiv:1409.3525
\bibitem{Y09}
	H.~P.~Yuen, 
	``Key generation: foundations and a new quantum approach,"
	IEEE J. Sel. Top. Quantum Electron., vol.15, no.6, pp.1630-1645, 2009.
\bibitem{Y10}
	H.~P.~Yuen, 
	``Fundamental quantitative security in quantum key generation,"
	Phys. Rev. A, vol.82, no.6, 062304, 2010.
\bibitem{Y13b}
	H.~P.~Yuen,
	``Essential elements lacking in security proofs for quantum key distribution."
	Proc. SPIE, vol.8899, 88990J, 2013.
\bibitem{Y13c}
	H.~P.~Yuen, ``On the foundations of quantum key distribution --- Reply to Renner and beyond," 
	Tamagawa University Quantum ICT Research Institute Bulletin, vol.3, no.1, pp.1-8, 2013; available at {\tt http://www.tamagawa.jp/research/quantum/
	bulletin/pdf/Tamagawa.Vol.3-1.pdf}
\bibitem{Y13d}
	H.~P.~Yuen, 
	``On the nature and claims of quantum key distribution (QKD),"
	Lecture at Tamagawa University, 5 Dec 2013; 
	available at {\tt http://www.tamagawa.jp/research/
	quantum/openlecture/}
\bibitem{Hirota12}
	O.~Hirota, 
	``Incompleteness and limit of quantum key distribution theory 
	- Yuen theory vs Renner theory -,"
	Tamagawa University Quantum ICT Research Institute Bulletin, 
	vol.2, no.1, pp.25-34, 2012;
	available at {\tt http://www.tamagawa.jp/research/
	quantum/bulletin/pdf/Tamagawa.Vol.2-6.pdf}
\bibitem{Hirota13}
	O.~Hirota, 
	``Misconception in theory of quantum key distribution - Reply to Renner -,"
	 quant-ph arXiv:1306.1277v1
\bibitem{Hirota14}
	O.~Hirota, 
	``A correct security evaluation of quantum key distribution,"
	quant-ph arXiv:1409.5991v1

\bibitem{L92}
	T.~Lindvall, 
	{Lectures On The Coupling Method}, 
	Wiley\&Sons 1992.
\bibitem{T00}
	H.~Thorisson, 
	Coupling, Stationarity, and Regeneration, 
	Springer 2000.
\bibitem{LPW09}
	D.~A.~Levin, Y.~Peres, and E.~L.~Wilmer, 
	Markov Chains and Mixing Times, 
	AMS 2009.
\bibitem{NC10}
	M.~A.~Nielsen, and I.~L.~Chuang, Quantum Computation and Quantum Information, 10th anniversary edition, Cambridge University Press 2010.

\bibitem{Shannon49}
	C.~E.~Shannon,
	``Communication theory of secrecy systems," 
	Bell System Technical Journal, vol.28, no.4, pp.656-715, 1949.
\end{thebibliography}
\end{document}